\documentstyle[aps,prl,preprint]{revtex}
\begin{document}
\bibliographystyle{unsrt}

\draft
\title{Energy thresholds for discrete breathers in
one-, two- and three-dimensional lattices}
\author{S. Flach$^1$, K. Kladko$^1$ and R. S. MacKay$^2$}
\address{$^1$ Max-Planck-Institute for Physics of Complex Systems, Bayreuther
Str. 40 H.16, D-01187 Dresden, Germany \\
$^2$ Department of Applied Mathematics and Theoretical Physics,
Silver Street, University of Cambridge, CB3 9EW England}
\date{\today}
\maketitle
\begin{abstract}
Discrete breathers are time-periodic, spatially localized solutions
of equations of motion for
classical degrees of freedom interacting on a lattice.  They
come in one-parameter families.
We report on studies of
energy properties of breather families in one-, two- and three-dimensional
lattices. We show that breather energies have a positive lower bound
if the lattice dimension of a given nonlinear lattice is greater
than or equal to a certain critical value.
These findings could be important for the experimental
detection of discrete breathers.
\end{abstract}


\newpage

Recently progress has been achieved
in the understanding of localized excitations in nonlinear lattices.
Discrete breathers (DBs)
are time-periodic, spatially localized solutions of
equations of motion for
classical degrees of freedom interacting on a lattice 
\cite{kk74},\cite{st88},\cite{tks88}.
Nowadays it is known that the reason
for the generic existence of DBs is the {\sl discreteness of the
system} paired with the {\sl nonlinearity} of the differential equations
defining the evolution of the system \cite{cp90},\cite{st92-jpsj-2}.
Thus one can avoid
resonances of multiples of the discrete breather's frequency $\Omega_b$
with the phonon spectrum $\Omega_q$ of the system \cite{sf94}.
If the coupling is weak the phonon spectrum consists of narrow bands.
The nonlinearity and the narrowness of the phonon bands
allows for periodic orbits whose frequency
and all its harmonics lie outside the phonon spectrum.
For some classes of system, existence proofs of breather solutions
have been published \cite{ma94},\cite{sf95-pre-1},\cite{db96}.
A list of references is given in \cite{wwwlist}.

For generic Hamiltonian systems, periodic orbits occur in one-parameter
families, and discrete breathers are no exception.  In many cases, the
energy can be used as parameter along the family, but as is well known,
the energy can have turning points along a family of periodic orbits.
Mathematically, such a turning point in energy is called a {\it saddle-centre
periodic orbit}.

The main message of this paper is that in 3D lattices, a
turning point (in fact, minimum) in energy is almost inevitable for
discrete breather families.

One important property of DBs is their generic existence for weak enough
coupling,
independent of the lattice dimension \cite{st92},\cite{ma94}.
This means that DBs are not just a
1D curiosity but could be interesting
from the point of view of applications. The experimental detection of DBs
requires some additional knowledge about their properties.
In this contribution we give heuristic arguments
that the energy of a DB family
has a positive lower bound for lattice dimension $d$ greater than or equal to
some $d_c$, whereas
for $d<d_c$ the energy goes to zero as the amplitude goes to zero,
and we confirm these predictions numerically.
The critical dimension $d_c$ depends on details of the system but
is typically $2$ and never greater than $2$.
Furthermore, for $d>d_c$, the minimum in energy
occurs at positive amplitude and finite localisation length.
Consequently experiments could be designed to look for activation
energy thresholds for localized excitations.

Let us consider a $d$-dimensional hypercubic lattice with $N$ sites.
Each site is labeled by a $d$-dimensional vector $l \in Z^d$.
Assign to each lattice site a state $X_l \in R^f$, where $f$
is the number of components and is to be finite.
The evolution of the system is assumed to be given by a Hamiltonian of
the form
\begin{equation}
H = \sum_l{H_{loc}(X_l) + H_{int}(X_l,\{X_{l+s}\})}, \label{1}
\end{equation}
where $H_{int}$ depends on the state at site $l$ and the states $X_{l+s}$ in a
neighbourhood.  We assume that $H$ has an equilibrium point at
$X_l=0$, with $H(\{X_l=0\}) = 0$.

DB solutions come in one-parameter families. The parameter can be
the amplitude (measured at the site with maximum amplitude),
the energy $E$ or the breather
frequency $\Omega_b$. It is anticipated (and was found both
numerically and through some reasonable approximations \cite{kk74}) that
the amplitude can be lowered to arbitrarily small values,
at least for some of the families for an infinite lattice.
In this zero amplitude limit, the DB frequency
$\Omega_b$ approaches an edge of the phonon spectrum $\Omega_q$.
This happens because the nonresonance condition $\Omega_q / \Omega_b
\neq 0,1,2,3,...$ has to hold for all solutions of a generic DB family
\cite{sf94}.
In the limit of zero amplitude, the solutions
have to approach solutions of the linearized equations of motion,
thus the frequency $\Omega_b$ has to approach some $\Omega_q$,
but at the same time not to coincide with any phonon frequency.
This is possible only if the breather's frequency tends
to an edge $\Omega_E$ of the phonon spectrum in the limit of zero breather
amplitude.
If we consider the family of nonlinear plane waves which
yields the corresponding band edge plane wave in the limit of zero
amplitude $A$, then its frequency $\Omega$  will depend on $A$ like
\begin{equation}
|\Omega -  \Omega_E| \sim A^z \label{z}
\end{equation}
for small $A$, where the ``detuning exponent" $z$ depends on the type of
nonlinearity of the Hamiltonian (\ref{1}), and
can be calculated using standard perturbation theory \cite{Naifeh}.

It is tempting to check then whether
the breather appears through a bifurcation
from a periodic orbit which is a normal mode of the linearized
equations of motion for any system with finite $N$.
Band edge plane waves of the linearized
equations of motion can be continued to non-zero amplitudes for
the general nonlinear system. The stability analysis of
these periodic orbits yields the possibility of tangent bifurcations
(collision of Floquet multipliers at +1) if some algebraic inequalities
of the expansion coefficients of $H$ in (\ref{1}) are met
\cite{sf96}. It has been also shown that the orbits which bifurcate
from the plane wave are not invariant under discrete translations
and have the shape of discrete breathers \cite{sf96}. It has been
conjectured that the new bifurcating orbits {\sl are} discrete
breathers. Subsequently it was successfully explained why discrete
breathers exist or not for certain models by
analyzing the above-mentioned algebraic inequalities \cite{sf96}. Numerical
studies confirm these findings \cite{sp94} for some one-dimensional models.

The above-mentioned analysis of stability of band edge plane waves 
was carried out for systems with detuning exponent $z=2$ and large $N$.
The critical amplitude $A_c$ of the plane waves at the bifurcation point
depends on the number of lattice sites as $A_c \sim N^{-1/d}$ \cite{sf96}.
We see that the amplitudes of the new orbits bifurcating from the plane
wave become small in the limit of large system size.
If the energy of the system is given by a positive definite quadratic form
in the variables $X$ in the limit of small values of $X$ it follows
for the critical energy of the plane wave at the bifurcation
point \cite{sf96}
\begin{equation}
E_c \sim N^{1-2/d} \;\;. \label{4}
\end{equation}
Result (\ref{4}) is surprising, since it predicts that for
$z=2$ the energy of a DB for small amplitudes
should diverge for an infinite lattice with $d=3$ and stay finite (nonzero)
for $d=2$, whereas if $d=1$ the  breather energy will tend
to zero (as initially expected) in the limit of small amplitudes
and large system size. The whole construction depends on the
validity of the assumption that the new periodic orbits bifurcating
from the plane wave through the above-mentioned tangent bifurcation are
indeed DBs.

It is not known how to prove this assumption. But we can estimate
the discrete breather energy in the limit of small amplitudes and
compare the result with (\ref{4}). 
Define the amplitude of a DB to be the largest of the amplitudes of
the oscillations over the lattice.  Denote it by $A_0$ where we define
the site $l=0$ to be the one with the largest amplitude. The
amplitudes decay in space away from the breather center,
and by linearising about the equilibrium state and making a continuum
approximation, the decay is found to be given by
$A_l \sim C F_d(|l|\delta)$
for $|l|$ large, where $F_d$ is a dimension-dependent function
\begin{eqnarray}
F_1(x) = e^{-x} \;\;,\;\;F_3(x) = \frac{1}{x}e^{-x}\;\;\label{f13}
\\
F_2(x) = \int \frac{e^{-x\sqrt{1+\zeta^2}}}{\sqrt{1+\zeta^2}} d\zeta
\;\;, \label{f2}
\end{eqnarray}
$\delta$ is a spatial decay exponent to be discussed shortly, and $C$ is
a constant which we shall assume can be taken of order $A_0$.
To estimate the dependence of the spatial decay exponent $\delta$ on
the frequency of the time-periodic motion $\Omega_b$ (which is
close to the edge of the linear spectrum) it is enough to
consider the dependence of the frequency of the phonon spectrum $\Omega_q$
on the wave vector $q$ when close to the edge. Generically this
dependence is quadratic $(\Omega_E-\Omega_q) \sim |q-q_E|^2$ where $\Omega_E
\neq 0$ marks the frequency of the edge of the linear spectrum
and $q_E$ is the corresponding edge wave vector.
Then analytical continuation of $(q-q_E)$ to ${\rm i} (q-q_E)$ yields a
quadratic dependence $|\Omega_b - \Omega_E| \sim \delta^2$.
Finally we must insert the way that the detuning of the breather frequency from
the edge of the linear spectrum $|\Omega_b - \Omega_E|$ depends on
the small breather amplitude.  Assuming that the
the weakly localized breather frequency detunes with amplitude as
the weakly nonlinear band edge plane wave frequency
this is
$|\Omega_b - \Omega_E| \sim A_0^z$.
Then $\delta \sim A_0^{z/2}$.

Now we are able to calculate the scaling of the energy of the discrete breather
as its amplitude goes to zero by
replacing the sum over the lattice sites by an integral
\begin{equation}
E_b \sim \frac{1}{2} C^2 \int r^{d-1} F_d^2(\delta r) {\rm d}r
\sim A_0^{(4-zd)/2}\;\;. \label{5}
\end{equation}
This is possible if the breather persists for small amplitudes
and is slowly varying in space.
We find that if $d > d_c=4/z$ the breather energy diverges
for small amplitudes, whereas for $d < d_c$ the DB energy tends
to zero with the amplitude. Inserting $z=2$ we obtain $d_c=2$,
which is in accord with the exact results on the plane wave stability
\cite{sf96} and thus strengthens the conjecture that discrete breathers
bifurcate through tangent bifurcations from band edge plane waves.
Note that for $d=d_c$ logarithmic corrections may apply to (\ref{5}), 
which can lead
to additional variations of the energy for small amplitudes.

An immediate consequence is that if $d \geq d_c$, the energy of a breather
is bounded away from zero.
This is because for any non-zero amplitude the breather energy
can not be zero, and as the amplitude goes to zero the energy goes to a
positive limit ($d=d_c$) or diverges ($d>d_c$).
Thus we obtain an energy threshold for
the creation of DBs for $d \geq d_c $. This new
energy scale is set by combinations of the expansion coefficients
in (\ref{1}).
If $z=2$ with $|\Omega-\Omega_E| \sim \beta A^2$ for the nonlinear
plane waves, and the energy per oscillator $E \sim gA^2$ and the spatial
decay exponent $\delta$ is related by $|\Omega_b-\Omega_E| \sim \kappa
\delta^2$, then the energy threshold $E_{min}$ is of the order of $\kappa
g/\beta$, and the minimum energy breather in 3D has spatial size of the
order of the lattice spacing, independently of $\kappa, g$ and $\beta$.
One should allow for a factor of $(2+d)$ for underestimating the true
height of the minimum and the contributions of nearest neighbours.

To confirm our findings, we performed
numerical calculations.
First we study the discrete nonlinear Schr\"odinger (DNLS) equation
\begin{equation}
\dot{\Psi}_l = {\rm i}(\Psi_l + |\Psi_l|^{\mu-1}\Psi_l
+ C\sum_{m\in N_l}\Psi_m),
\end{equation}
where $N_l$ denotes the set of nearest neighbours of $l$.  The detuning
exponent $z$ is easily seen to be $\mu -1$.
Making the substitution $\Psi_l = A_l {\rm e}^{{\rm i} \Omega_b t}$
we solve the algebraic equations for the real amplitudes
$A_l$. Numerically this is implemented by considering the case of
large breather amplitude $A_0$ first. Then the breather is essentially
given by $A_0 \approx (\Omega_b -1)^{1/(\mu -1)}$ and $A_{l \neq 0} =0$.
Next we define a functional $G$ which is the sum over the squares
of differences between left hand and right hand parts of all algebraic
equations for the amplitudes. This functional is minimized by
gradient descent, where the initial guess is the large amplitude
approximate solution. Finally the frequency $\Omega_b$ is varied in small
steps and the breather solution is traced. In Fig.1 we show the
resulting breather energy as a function of the amplitude $A_0$
for $\mu=3$ and $d=1,2,3$. The results are in full accord with the
predictions. For $d=3$ the above estimate of the minimum energy 
yields a value of 0.2 with $\beta=g=1$ and $\kappa=C=0.1$. The mentioned
factor $(2+d)=5$ accounts for the deviation from the true value of 1. 
Fig.2 shows the amplitude distribution of the discrete breather
with minimum energy in the $(x,y)$-plane crossing
the breather center for $d=3$. The minimum energy breather is
strongly localized - its spatial width is only few lattice spacings.
In Fig.3 we show results for $d=1$ and $\mu=3,5,7$.
Again we find full agreement. Note that even one-dimensional lattices
exhibit positive lower bounds on breather energies if $\mu \geq 5$.
This $d=1$ result has also been predicted using variational techniques
\cite{mw96}.

To demonstrate that the numerical results are not an artefact of
the DNLS case, we study the $d$-dimensional
nonlinear Klein-Gordon lattice
\begin{equation}
\ddot{U}_l = -U_l - U_l^{\mu} - C\sum_{m \in N_l}(U_l-U_m).
\end{equation}
The detuning exponent $z$ is given by
$\mu -1$ for $\mu$ odd and $2\mu - 2$ for $\mu$ even.
Again the discrete breather with large amplitude
is essentially an on-site excitation and given by
$\ddot{U}_0 = -U_0 - U_0^{\mu}$ and $U_{l\neq 0} =0$. The equations
of motion are integrated numerically for a given set of initial
conditions $\{U_l(t=0),\dot{U}_l(t=0)\}$ over the breather period
$T_b=2\pi/\Omega_b$. The functional $G=\sum_l\left( (U_l(T_b)-U_l(0))^2
+(\dot{U}_l(T_b)-\dot{U}_l(0))^2 \right)$ is minimized
with respect to the initial conditions using gradient descent.
This method allows us to perform a reliable numerical
calculation of DBs in $3$-dimensional arbitrary lattices.
The result in Fig.4 for $\mu=3$ and $d=3$
is again in full accord with the predictions.

We can predict that a modified DNLS system with an additional
term $v_{\mu'}|\Psi_l|^{\mu' -1}\Psi_l$ can exhibit complex
curves $E_b(A_0)$. For example, for $d=1$, $\mu=7$, $\mu'=3$ and
$v_{\mu'}=0.1$, the $E_b(A_0)$-dependence will be nearly
identical to the case $v_{\mu'}=0$ already considered, if the
amplitude $A_0$ is not too small.
Then $E_b(A_0)$ will show
a minimum at a non-zero value of $A_0$. For small $A_0$ however
the energy of the breather will ultimately decay to zero, so
the curve has a maximum for smaller amplitudes! The dashed line in
Fig.3 shows the numerical calculation, which coincides with our prediction.

Our findings should help to detect discrete breathers in experimental
realizations like the dynamics of atoms in crystals. For
a $3$-dimensional crystal we predict a positive energy threshold
for the excitation of discrete breathers.

Another consequence of our work is that breather solutions
belonging to parts of the family where the energy is decreasing with increasing
amplitude are dynamically unstable, whereas those in
the other parts have a good chance of being dynamically stable.
This can be seen from a Poincar\'e
map of the phase space flow around the breather orbits.
The minimum energy breathers correspond to saddle-centre
bifurcations, since no breather solutions will exist if the energy
is lowered beyond the minimum breather energy.

A similar phenomenon occurs in polaron theory. In a three-dimensional
lattice, two polarons of unit electric charges exist above
a certain parameter threshold (large and small polaron) \cite{de95}.

Summarizing, we have shown that discrete breather families have positive
lower energy bounds if the dimension of the lattice is larger or
equal to some critical value which in turn is defined by the
power of the first nonlinear expansion term in the equations
of motion. These results are expected to be of importance for the experimental
detection of discrete breathers, because the minimum energy of a breather
family should show up as an activation energy.

\newpage

FIGURE CAPTIONS
\\
\\
Fig.1:
\\
Breather energy versus  amplitude for the DNLS system
in one, two and three lattice dimensions.
Parameters $C=0.1$ and $\mu=3$ for all cases.
System sizes for $d=1,2,3$: $N$=100, $N$=$25^2$, $N$=$31^3$, respectively.
The estimated points ($A;E$) of bifurcation of the band edge plane wave
for $d=1,2,3$ are:
(0.014;0.024), (0.064;5.53), (0.097; 237), respectively.
\\
\\
Fig.2:
\\
Amplitude distribution of the minimum energy breather solution
of the DNLS system with $d=3$,
$\mu=3$, $C=0.1$
and $N=31^3$. Actually only a distribution in a cutting $(x;y)$ plane
is shown (the plane cuts the center of the breather). The intersections
of the grid lines correspond to the actual amplitudes, the rest of
the grid lines are guides to the eye.
\\
\\
Fig.3:
\\
Breather energy versus maximum amplitude for the DNLS
system in one lattice dimension and for
three different exponents $\mu=3,5,7$ (solid lines). The system size is
$N=100$ and the parameter $C=0.1$.
The dashed line is for the modified system (cf. text).
\\
\\
Fig.4:
\\
Breather energy $E_b$ versus frequency detuning $(\Omega - \Omega_E)$
for a 3D Klein-Gordon lattice.
Parameters $\mu=3$ and $C=0.1$.
System size $N$=$10^3$.

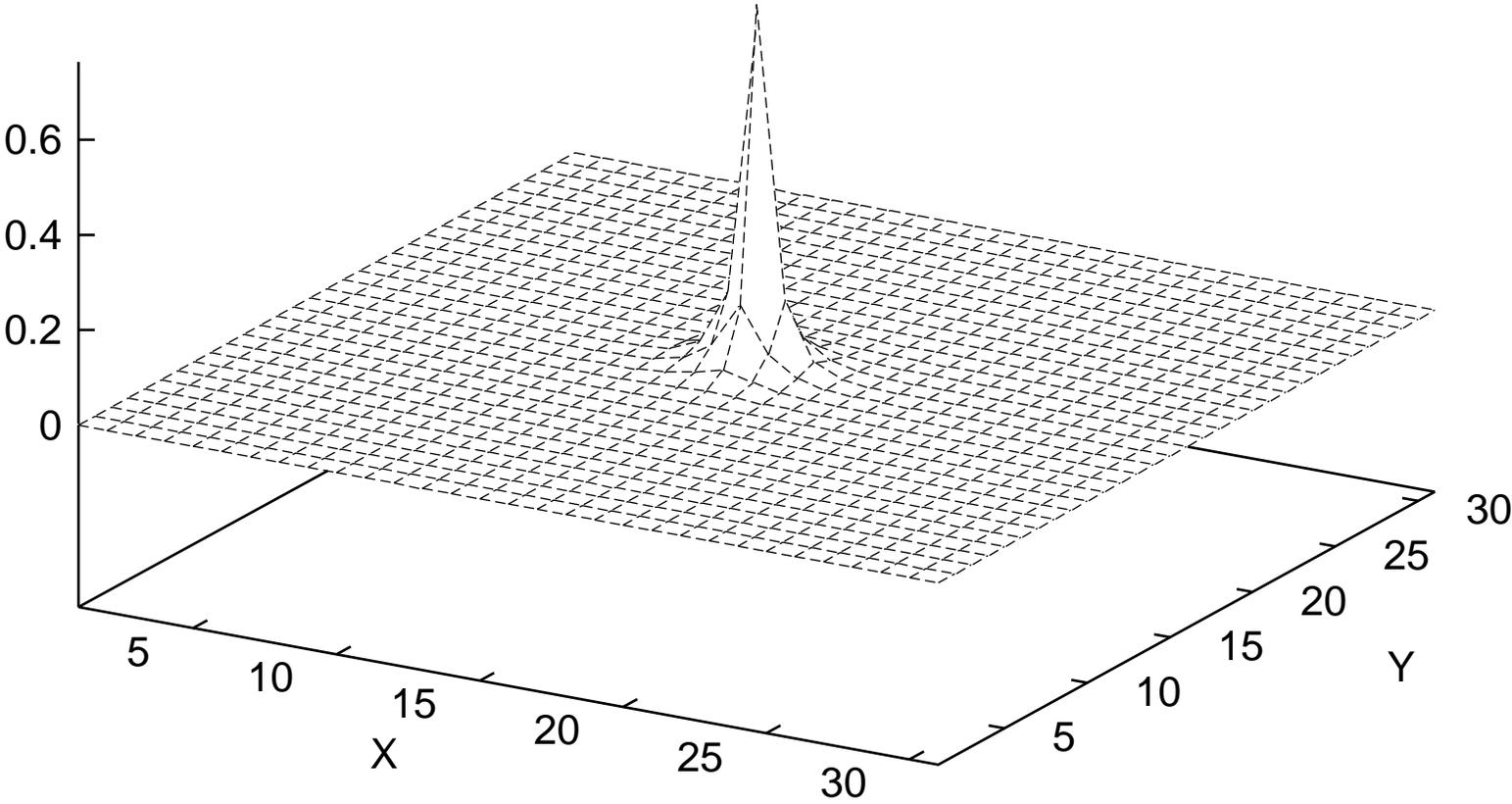
\end{document}